\documentclass[twocolumn,showpacs,preprintnumbers,amsmath,amssymb]{revtex4}
\usepackage{graphicx}
\usepackage{dcolumn}
\usepackage{bm}

\newcommand{\bi}{\begin{itemize}}
\newcommand{\ei}{\end{itemize}}
\newcommand{\be}{\begin{equation}}
\newcommand{\ee}{\end{equation}}
\newcommand{\ba}{\begin{eqnarray}}
\newcommand{\ea}{\end{eqnarray}}
\newcommand{\bse}{\begin{subequations}}
\newcommand{\ese}{\end{subequations}}

\begin{document}


\title{Theoretical Value of the Electromagnetic Coupling Constant} 

\author{Yair Goldin$^\dagger$}

\affiliation{$^\dagger$Facultad de Ciencias,\\ Universidad Nacional Aut\'onoma de
M\'exico,\\ Circuito Exterior, C. U., \\M\'exico D.F.,  04510, M\'exico}

\email{yair@nuclecu.unam.mx}

\altaffiliation[Mail correspondence to\,]{\,\,Lomas de Tarango 155, M\'exico DF 01620,
M\'exico}

\begin{abstract}
An insight into bispinor analysis makes it possible to describe  the 
electron in selfaction as a fundamental 
steady state. The electromagnetic theory, and
the Dirac equation for the study of an electron in 
presence  of external
potentials, follow as natural extensions of the 
equations that rule the electron in
selfaction.  The electromagnetic coupling constant ($\alpha$) 
and the coupling constant ($\beta$) of a gauge invariant matrix vector potential 
 are interrelated by the equation that defines the electron structure. Here, 
two bispinor components carry $1/3$ and $2/3$ of the physical properties of the electron: electric charge, mass,
spin and magnetic moment. These fractions of the electron charge seem to be a feature common to both leptons and
hadrons.  An eigenvalue equation involving the invariants of the 
selfpotentials ultimately determines $\alpha$ and $\beta$.
\end{abstract}


\maketitle
 
\section{introduction}
 Free particles, by definition,  are not subject to experiment or observation; nonetheless, it is of
episthemological interest to search for the principles that make their existence
possible and to find out the relation of these principles to the 
physics of observable phenomena. First,
 the kinematic behavior of free particles is ruled by the principle of inertia; 
otherwise the parameter of relative velocity at the core of the Lorentz transformations
could not be defined and covariance would lose significance. Indeed, two inertial systems
of reference in relative motion {\textit{with a particle affixed to one of them}} compose the
scenario in which the concept of relative velocity acquires physical meaning. In Quantum Mechanics 
(QM),  the radial potential in the solution of the hydrogen atom implies a proton at the origin of
an inertial system of reference, more precisely, it is  the center of mass of the atom which is in
fixed position: the principle of inertia is an objective physical reality, perhaps the same one
that Einstein used to refer to. So, why is it  that the wave function cannot represent a free
electron in fixed position?
 
Soon after the advent of the Dirac equation physicists tried to complete the study of the free electron,  
and in order to do so selfaction had to be reckoned with.  By the late 40's the latest techniques of QED
were applied to the case in question~\cite{[1],[2]}. Although, by emitting and absorbing
virtual photons along its free path, the electron selfenergy separates into two parts: a
bare mass whose origin can not be explained, and an electromagnetic mass which cannot be
made finite.

The unsuccessful attempts of QED to take into account self action in the
propagator of the Dirac equation,
	\begin{equation}  
	\gamma_\mu\,\partial_\mu\,\psi \ = \
	m\,c^2\,\gamma_{_5}\,\psi,\label{eq1}
	\end{equation}
led the study of the free electron to a dead end.
The Dirac equation is useless to describe the electron \textit{per se}. Moreover, it is not 
possible  to say whether or not its solutions, plain waves uniformly exteding throughout space,  
are meaningful.  

Considering that a free electron  
can  be ``at rest'', that selfaction is 
the only cause of the electron mass, that the electron 
in isolation has definite energy
(steady state), and that the positron has positive energy, 
 suggests the 
fundamental equation,
	\begin{equation}
	\gamma _\mu \left[ {\partial _\mu -S_\mu } \right]\,\psi
	\ = \ 0,\label{eq2}
	\end{equation}
where the Coulomb potential is part of the selfaction $S_\mu$.

The natural extension of eq. (2)  
for the study of an electron under action of external
potentials would be,
	\begin{equation}  
	\gamma_\mu\,\left[\partial_\mu
	-e\,A_\mu\right]\,\psi \ = \
	m\,c^2\,\gamma_{_5}\,\psi,\label{eq3}
	\end{equation}
since the interaction term in both equations can 
shift the original energy at their right
hand side to the level of energy of a steady state. 

The natural extension of eq(3) for the solution of potential problems where the
electron is free in a well defined region of spacetime is eq.(1). However, boundary conditions on
its positive energy solutions  are necessary to endow 
the latter  with physical
meaning.  The role 
of QM in the study of the free electron is to enhance the
principle of inertia and to answer the larger 
question: how can the electron possibly
exist?

\section{The selfpotentials}

The Electric charge, as it is  understood at the present time, plays a
dual role in the electromagnetic theory: it is the source of the potentials and the recipient 
of their action. In the 
classical expression of density of electrostatic energy $\rho\,\phi $, the charge
density $\rho$ is the recipient of the action of $\phi$. When $\rho$ is identified with
the Laplacian of $\phi$, the electrostatic energy, localized where $\rho$ and $\phi$
coexist, all of a sudden disperses throughout space  as the electrostatic field
squared. Such ambiguity and the fact that a point charge at 
the singularity of the potential yields divergent selfenergy suggest to take a second look
 on Maxwell's concept of electric charge.
 
 If
the Dirac $\delta$ is assumed to be 
source of the radial field $r^{-2}\,\hat{\textrm{r}}$ ,
we get the contradiction $0=1 $ when the space integral of Gauss law ($\nabla \cdot \vec{\textrm{E}}  
=e\,\delta$) is carried out.
	\ba 
	\int_{\textrm{space}}{ \nabla \cdot \vec{ \textrm{E} }}=
	4\pi \int_0^\infty{\partial _r(1)}\,dr \ne e.\label{eq4}
	\ea
Actually, the equation
	\ba \displaystyle &\lim_{\beta\to 0}& 
	\frac{4\pi}{e}\,\int_{\beta^2(e^2/mc^2)}^\epsilon{\nabla \cdot
	\vec{\textrm{E}}\,r^2\,dr}=1,
	\ea 
where  $\epsilon$ is any radius, $\beta$ is a numerical parameter and $e^2/(m c^2)$ is the classical
electron radius, implies the vanishing of the electrostatic field at
$r=0$. Consider a radial function $\Delta$ equal to $1$ if $r\ne0$ and equal to $0$ if
$r= 0$, with a continuos representation with mathematical meaning only under the integral
sign: 
\begin{equation}\textrm{rep}(\Delta) = (\displaystyle\lim_{\beta\to 0};\,\exp
[-\beta(e^2/mc^2)r^{-1}]).\nonumber\end{equation} 
Substituting the electrostatic field
\begin{equation}\vec{\textrm{E}}=(4\pi)^{-1}\,e\,\exp
[-\beta(e^2/mc^2)r^{-1}]\,r^{-2}\,\hat{\textrm{r}},\end{equation} 
in eq.(5) we get
\ba\displaystyle &\lim_{\beta\to 0}&
\int_{\beta^2(e^2/mc^2)}^\epsilon{\partial_r\left(\,\exp [-\beta(e^2/mc^2)\,r^{-1}]
\right)\,dr} \nonumber \\= \displaystyle &\lim_{\beta\to 0}&\left(\exp
[-\beta(e^2/mc^2)\,\epsilon^{-1}]-\exp\left[-\beta^{-1}\right]\right)=1.\nonumber\\
\ea
The result above is hardly surprising since the weak derivative of the Heaviside
step function ($f(x) =0$ if $x\leq 0$, and $f(x) =1$ if $x> 0$), is identified with a
functional 
\begin{equation}\int_{-\infty}^\infty\,\frac{\partial f}{\partial
x}\,dx=\int_{-\infty}^\infty{\delta(x)\,dx} =1.\nonumber\end{equation}
However, the electric field of the electron does not go to zero at $r=0$. The fact that the
gradient of the potential $r^{-1}$ diverges at $r=0$ is precisely the reason which makes it
impossible to prove its source--dependency.   

The vanishing of the Laplacian of the inverse distance is a fundamental property that
distinguishes physical space from mathematical spaces since the equation below 
holds only for $n=3$
\begin{equation}\left[\frac{\partial^2}{\partial x_1^2}+\frac{\partial^2}{\partial
x_2^2}+\dots+\frac{\partial^2}{\partial
x_n^2}\right]\,\left[x_1^2+x_2^2+\dots+x_n^2\right]^{-1/2}=0. \end{equation}

It is important to note that  admitting the Coulomb potential as a solution of
 a differential equation implies admitting  infinity as an essential concept ($r^{-1}$ diverges at 
$r=0$). Therefore, there is no reason not to admit that the term $-r^{-2}$ is the faithful
representation of the slope of $r^{-1}$ throughout space. The assertion above takes us back to
statement (4), which simply shows that the  Coulomb potential is much too fundamental 
to force on it a functional dependence on a source.  

Equations (2) and (8) are corner stones of a unified theory of steady states. 
In familiar terms, the electron is to be conceived as a  recipient point (or recipient charge) 
 being acted upon by a repulsive Coulomb potential. 
The electric charge is not the
source, but merely the coefficient of the radial function $r^{-1}$. 
On the other hand,
the position of the recipient point  is given by the probability
distribution associated with the wave functions. 
It will be shown that the probability
distribution and its derivatives 
up to any order vanish at $r=0$, thus confirming
the absence of tangible sources at the singularity. 
The distribution reaches a peak at
about 4 millionths of the classical electron radius, 
rapidly decays and vanishes in a
smooth manner at the radius just mentioned.

Electromagnetic signals from the singularity are necessarily related to the 4-vector 
\begin{equation}r_\mu=(r,
\vec{\textrm{r}}),\end{equation} 
of null length. When $r_\mu$ is contracted with the 4--velocity of the singular point, 
\begin{equation} u_\mu \ = \
(1,\vec{\textrm{u}})\,(1-\textrm{u}^2)^{-1/2}\ \to\,
(1,\vec{\textrm{0}}),\label{eq6}\end{equation}
(the arrow reads ``for a particle at rest reduces to'') yields the fundamental invariant,
\begin{equation} \textrm{I}_0  = \
(r_\mu\,u_\mu)^{-1}\,\to\,r^{-1}.\label{eq7}\end{equation}
This invariant allows for the construction of two 4--vectors: the Lienard--Wiechert
potentials 
\begin{equation} A_\mu = I_0\,u_\mu, \end{equation}
and the gauge invariant 
\begin{equation} B_\mu = \nabla_\mu\,I_0\quad \to \quad (0,r^{-2}\,\hat{\textrm{r}}).
\end{equation}

As a  direct consequence of 
their genesis, the potentials $A_\mu$ satisfy the fundamental  differential
equations
\ba  \nabla_\mu\,\nabla_\mu\,A_{\nu} \ = \
0,\label{eq9}\\ \nabla_\mu\,A_\mu \ = \ 0,\label{eq10}\ea
although, from a conceptual viewpoint, radial solutions 
of equation (14) are physical if and only if there are 
bispinor densities surrounding the singularities.

In order to write down in a concise manner the potentials of a continuous
distribution of electrons in arbitrary motion, one 
superposes the contribution of
each electron with the customary equations,
$\nabla_\mu\,\nabla_\mu\,A_{\nu}  = 
J_\nu$ and $\nabla_\mu\,A_\mu=0$. However, in this 
case $J_\nu$  would represent the
piecewise differentiable functions best 
approximating the current  singularities. The
vanishing of the 4--divergence\, $\nabla_\nu\,J_\nu$ \, states the
conservation of the net number of singular points.

Concerning the potential $B_\mu$, it is 
important to note the following: in Classical
Electromagnetism the vector potential of a point magnet is $\vec{\mu}\,\times\,
\vec{\textrm{r}}/r^3$; its quantum version would be proportional to
$\vec{\sigma}\,\times\, \vec{\textrm{r}}/r^3$, where the $\sigma$'s are the $4\times4$
block diagonal Pauli spin matrices ($\sigma_x,\,\sigma_y,\,\sigma_z$)
\ba \left[\begin{array}{llll} 
0&1&0&0\\
1&0&0&0\\
0&0&0&1\\
0&0&1&0\end{array}\right],\quad \left[\begin{array}{llll} 
0&-i&0&\,\,\,0\\
i&\,\,\,0&0&\,\,\,0\\
0&\,\,\,0&0&-i\\
0&\,\,\,0&i&\,\,\,0\end{array}\right],\quad \left[\begin{array}{llll} 
1&\,\,\,\,0&0&\,\,\,\,0\\
0&-1&0&\,\,\,\,0\\
0&\,\,\,\,0&1&\,\,\,\,0\\
0&\,\,\,\,0&0&-1\end{array}\right],\nonumber
\\\ea 
when $\vec{\sigma}\,\times\, \vec{\textrm{r}}/r^3$ is inserted in equation (2), the Dirac
matrices 
\ba \gamma_t &=& \left[\begin{array}{llll} 
1&\,\,\,\,0&\,\,\,\,0&\,\,\,\,0\\
0&\,\,\,\,1&\,\,\,\,0&\,\,\,\,0\\
0&\,\,\,\,0&\,\,\,\,1&\,\,\,\,0\\
0&\,\,\,\,0&\,\,\,\,0&\,\,\,\,1\end{array}\right],\quad \gamma_x=\left[\begin{array}{llll} 
0&\,\,\,\,0&\,\,\,\,0&\,\,\,\,1\\
0&\,\,\,\,0&\,\,\,\,1&\,\,\,\,0\\
0&\,\,\,\,1&\,\,\,\,0&\,\,\,\,0\\
1&\,\,\,\,0&\,\,\,\,0&\,\,\,\,0\end{array}\right],\nonumber\\
\gamma_y&=&\left[\begin{array}{llll}  0&\,\,\,\,0&\,\,\,\,0&-i\\
0&\,\,\,\,0&\,\,\,\,i&\,\,\,\,0\\
0&-i&\,\,\,\,0&\,\,\,\,0\\
i&\,\,\,\,0&\,\,\,\,0&\,\,\,\,0\end{array}\right],\quad \gamma_z=\left[\begin{array}{llll} 
0&\,\,\,\,0&\,\,\,\,1&\,\,\,\,0\\
0&\,\,\,\,0&\,\,\,\,0&-1\\
1&\,\,\,\,0&\,\,\,\,0&\,\,\,\,0\\
0&-1&\,\,\,\,0&\,\,\,\,0\end{array}\right],\nonumber\\
\ea
convert the matrix vector potential into an ordinary radial imaginary vector
$(2\,i\,r^{-2}\hat{\textrm{r}})$, the 4--vector version 
of which is $B_\mu$. That is, we
could consider that $B_\mu$ is the vector potential of the electron

In accordance with all the previous considerations, 
equation (2) unfolds as 
%
\ba \left(\frac{E}{\hbar c}-\frac{\alpha}{r}\right)\,\psi_1
&-&i\,\left(\partial_x-i\,\partial_y\right)\,\psi_4
-i\,\partial_z\,\psi_3\nonumber
\\
&+&\frac{i\,\beta\,e^2}{mc^2\,r^3}\,\left[\left(x-i\,y\right)\,\psi_4+
z\,\psi_3\right]\ = \ 0,\nonumber
\\
\nonumber
\\
\left(\frac{E}{\hbar c}-\frac{\alpha}{r}\right)\,\psi_2
&-&i\,\left(\partial_x+i\,\partial_y\right)\,\psi_3 +i\,\partial_z\,\psi_4\nonumber
\\
&+&\frac{i\,\beta\,e^2}{mc^2\,r^3}\,\left[\left(x+i\,y\right)\,\psi_3-\,z\,\psi_4\right]
\ = \ 0,\nonumber
\\
\nonumber
\\
\left(\frac{E}{\hbar c}-\frac{\alpha}{r}\right)\,\psi_3
&-&i\,\left(\partial_x-i\,\partial_y\right)\,\psi_2-i\,\partial_z\,\psi_1\nonumber
\\
&+&\frac{i\,\beta\,e^2}{mc^2\,r^3}\,\left[\left(x-i\,y\right)\,\psi_2+
\,z\,\psi_1\right] \ = \ 0,\nonumber
\\
\nonumber
\\
\left(\frac{E}{\hbar c}-\frac{\alpha}{r}\right)\,\psi_4
&-&i\,\left(\partial_x+i\,\partial_y\right)\,\psi_1+i\,\partial_z\,\psi_2\nonumber
\\
&+&\frac{i\,\beta\,e^2}{mc^2\,r^3}\,\left[\left(x+i\,y\right)\,\psi_1-
z\,\psi_2\right] \ = \ 0.\nonumber\\
\ea 
%
The time dependence $\exp\,(-i\,E\,t\,\hbar^{-1})$ has been taken into account, $\beta$
is a dimensionless parameter measuring the strength of the coupling with the imaginary
radial vector potential, $\gamma_\mu
\partial_\mu=i\,\hbar\,c\,(\gamma_t\,\partial_{ct}-\gamma_x\,\partial_x-\gamma_y\,\partial_y-
\gamma_z\, \partial_z)$ and $\alpha=e^2/(\hbar\,c)$. 




\section{The solutions.}

Since the potentials are radial the solutions can be written as follows~\cite{[3]}

\noindent
First solution:
\ba \psi_1 &=& 
\left[\frac{j+1-m}{2(j+1)}\right]^{1/2}F\,Y_{j+\frac{1}{2},m-\frac{1}{2}} \to 
\frac{F\,Y_{1,0}}{\sqrt{3}},\\
-\psi_2 &=&\left[\frac{j+1+m}{2(j+1)}\right]^{1/2}F\,Y_{j+\frac{1}{2},m+\frac{1}{2}} 
\to \sqrt{\frac{2}{3}}\,F\,Y_{1,1},\nonumber\\
\psi_3 &=& i\left[\frac{j+m}{2j}\right]^{1/2}G\,Y_{j-\frac{1}{2},m-\frac{1}{2}}\to
i\,G\,Y_{0,0},\nonumber\\
\psi_4 &=& i\left[\frac{j-m}{2j}\right]^{1/2}G\,Y_{j-\frac{1}{2},m+\frac{1}{2}}\to
0.\nonumber
\ea
\noindent
Second solution:\\
\ba\psi_1 &=& i\left[\frac{j+m}{2j}\right]^{1/2}L
\,Y_{j-\frac{1}{2},m-\frac{1}{2}}
\to i\,L\,Y_{0,0},\\
\psi_2 &=& i\left[\frac{j-m}{2j}\right]^{1/2}L\,Y_{j-\frac{1}{2},m+\frac{1}{2}}\to
0,\nonumber\\
\psi_3 &=& 
\left[\frac{j+1-m}{2(j+1)}\right]^{1/2}K\,Y_{j+\frac{1}{2},m-\frac{1}{2}} \to 
\frac{K\,Y_{1,0}}{\sqrt{3}},\nonumber\\
-\psi_4 &=& \left[\frac{j+1+m}{2(j+1)}\right]^{1/2}K\,Y_{j+\frac{1}{2},m+\frac{1}{2}} 
\to \sqrt{\frac{2}{3}}\,K\,Y_{1,1}.\nonumber\\\nonumber
\ea

\noindent
The spherical harmonics are already normalized to one. The arrow indicates  that
the value of the total angular momentum in this context is $j=1/2$.
($m=1/2$ is an option). The substitution of the sum  of these solutions in 
equation (\ref{eq2}) yields two identical systems of differential equations, each
containing but one pair of radial functions, $(F,G)$ or $(K,L)$~\cite{[3]}: 
	\bse
	\label{eq15}
	\ba \partial_r
	F+\frac{2F}{r}-\frac{\beta\,e^2\,F}{m\,c^2\,r^2}\ 
	&=& \left[\frac{E}{\hbar
	c}-\frac{\alpha}{r}\right]\,G,\\
	-\partial_r G+\frac{\beta\,e^2\,G}{m\,c^2\,r^2}\
	&=& \left[\frac{E}{\hbar c}-\frac{\alpha}{r}\right]\,F,
	\ea
	\ese
	\bse\label{eq16}
	\ba \partial_r
	K+\frac{2K}{r}-\frac{\beta\,e^2\,K}{m\,c^2\,r^2}\ 
	=&&\left[\frac{E}{\hbar
	c}-\frac{\alpha}{r}\right]\,L,\\
	-\partial_r L+\frac{\beta\,e^2\,L}{m\,c^2\,r^2}\
	=&&\left[\frac{E}{\hbar c}-\frac{\alpha}{r}\right]\,K,
	\ea
	\ese
We can
set either pair of functions equal to zero. The difference between  solutions (19) and (20) will
show up when their corresponding bispinor densities are written down.

The substitutions
	\bse
	\label{eq17}
	\ba r &=&\frac{e^2}{E}\,s,\quad 
	\quad
	\gamma\ = \ \frac{E}{mc^2},\nonumber\\
	F &=&\tilde F\,\exp(-\beta\,\gamma\,s^{-1}),\\
	G &=&\tilde G\,\exp(-\beta\,\gamma\,s^{-1}),
	\ea
	\ese
reduce the system to            
\bse\label{eq18}\ba
s^{-2}\partial_s(s^2\,\tilde F) \ = \ (1-s^{-1})\,\alpha\,\tilde G,\\
-\partial_s\,\tilde G \ = \ (1-s^{-1})\,\alpha\,\tilde F.\ea\ese

The radius $s=1$ is a regular singularity  and the first independent solution can be
expressed as 
\ba\tilde G \ = \ 1+\tilde G_1 +\tilde G_2 +\dots,\\
\tilde F \ = \ \tilde F_0+\tilde F_1 +\tilde F_2 +\dots,\ea
where $\tilde G_0=1$ generates this solution.  
Increasing subindex number  
implies increasing powers of $\alpha^2$ as 
coefficient of the functions. Most notably, all
functions except $\tilde G_0$ smoothly vanish at $s=1$. 
The procedure to solve the system is iterative and
runs as shown. First we insert $\tilde G_0$ 
in equation (24a), integrate and adjust the
integration constant to make $\tilde F_0$ vanish at $s=1$, that is:  
\ba s^{-2}\,\partial_s\,(s^2\,\tilde F_0)
&=&(1-s^{-1})\,\alpha\,\tilde G_0 \nonumber\\
\qquad\Rightarrow \quad \tilde F_0&=&(\alpha/6)\,(s^{-2}-3+2\,s).\ea
Next we substitute $\tilde F_0$ into (24b) integrate and adjust 
the integration constant 
to make $\tilde G_1$ vanish at $s=1$, thus:
\ba -\partial_s\,\tilde G_1 &=& (1-s^{-1})\,\alpha\,\tilde F_0\nonumber\\
\Rightarrow\qquad \tilde G_1 &=&
-(\alpha^2/12)\,(s^{-2}-2\,s^{-1}+6\,\ln(s)\nonumber\\
&& +9-10\,s+2\,s^2).\ea
Following the same procedure we get
\ba s^{-2}\,\partial_s\,(s^2\,\tilde F_1)&=&(1-s^{-1})\,\alpha\,
\tilde G_1 \nonumber\\
\Rightarrow \quad \tilde F_1&=&(\alpha^3/12)\,[(31/15) s^{-2}+s^{-2}\,\ln s
-3\,s^{-1}\nonumber\\  &+&3\,\ln s+4-(17/3) s-2\,s\,\ln s
+3\,s^2\nonumber\\
&-&(2/5)\,s^3],\ea
\ba -\partial_s \,\tilde G_2 &=& (1-s^{-1})\,\alpha\,\tilde F_1\nonumber\\
\Rightarrow \quad \tilde G_2 &=& (\alpha^4/12)\,[-(77/60)\,s^{-2}-(1/2)\,
s^{-2}\,\ln s\nonumber\\
&+&(91/15)\,s^{-1}+s^{-1}\,\ln s - (14/3)\,s-5\,s\,\ln s\nonumber\\
&-&(35/12)+ 7\,\ln s+(23/6)\,s^2+s^2\,\ln s\nonumber\\
 &-&(17/15)\,s^3+(1/10)\,s^4+(3/2)\,(\ln s)^2].\ea 
The second independent solution is expressed as 
\ba \tilde f&=&s^{-2}+\tilde f_1 +\tilde f_2 +\dots,\\
\tilde g&=&\tilde g_0+\tilde g_1 +\tilde g_2
+\dots\,,\ea
here,  $\tilde f_0=s^{-2}$ is the generator of the
solution. As in the former case,  all the other functions and their first derivatives 
vanish at $s=1$,
so:
\ba -\partial_s \,\tilde g_0 &=& (1-s^{-1})\,\alpha\,s^{-2}\nonumber\\
\Rightarrow \quad \tilde g_0&=&-(\alpha/2)\,(1-s)^2\,s^{-2},\ea
\ba s^{-2}\,\partial_s\,(s^2\,\tilde f_1) &=
& (1-s^{-1})\,\alpha\,\tilde g_0\nonumber\\
\Rightarrow \quad \tilde f_1 &=& (\alpha^2/12)\,(11\,s^{-2}+6\,s^{-2}\,
\ln s
-18\,s^{-1},\nonumber\\
&+& 9 -2\,s),\ea
\ba -\partial_s\,\tilde g_1 &=&(1-s^{-1})\,\alpha\,\tilde f_1\nonumber\\
\Rightarrow \quad \tilde g_1 &=& (\alpha^3/12)\,(-7\,s^{-2}-3\,s^{-2}\,\ln s
+35\,s^{-1}\nonumber\\
&+&6\,s^{-1}\,\ln s -18 +27\,\ln s -11\,s +s^2).\nonumber\\
\ea
\begin{figure}
\centering
\includegraphics[height=15cm]{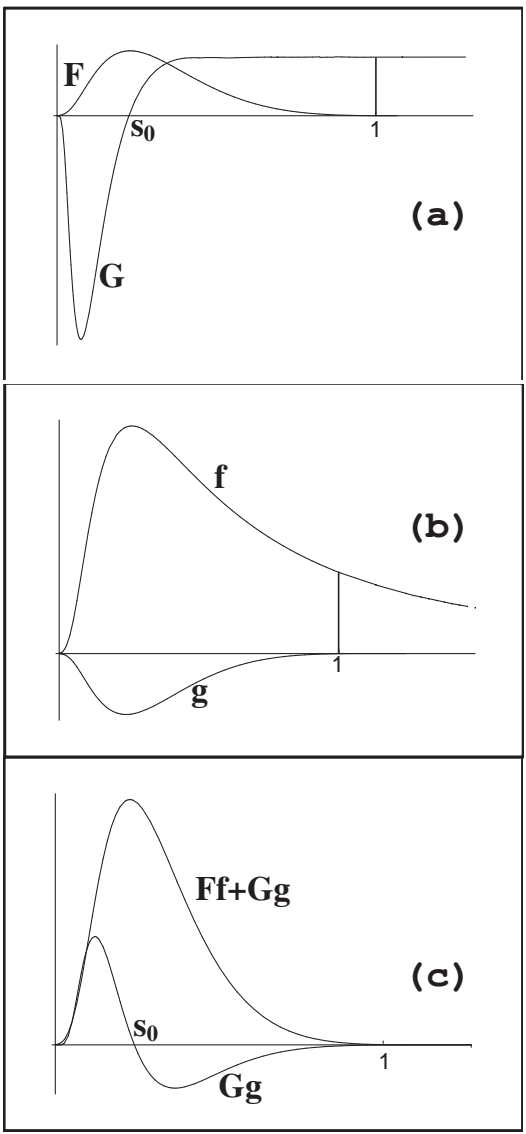}
%
%
\caption{The graphs  show the main features of the functions. All functions  vanish in
a smooth manner at the origin of coordinates. In the first independent solution, function
$G$ cuts the $s$--axis only once at  $s^2\approx\alpha^2/12 $. Functions $F$ and $g$ and their first derivative 
vanish 
 at $s=1$.   Beyond the classical electron radius, the product of the two
solutions is null. Thus, the energy of the electron is confined within its classical
radius.}
\label{fig:1}       
\end{figure}
 Since functions: $F,\, g,\,  g\, G$ and $ fF$
vanish in a smooth manner at 
$s=1$, the physical properties of the electron are to be represented with covariant
densities involving solely  the product of the two independent solutions:
\ba (2/\alpha)\,\tilde G\,\tilde g\,s^2
&=&
-(1-s)^2 \nonumber \\
&+& 
(\alpha^2/12)\,[s^{-2}-4\,s^{-1}+ 40\,s-5\,s^2\nonumber\\ 
&+&
60\,s^2\,\ln s
-36\,s^3+4\,s^4]\nonumber \\ 
&+&
(\alpha^4/12) [(147/60)\,s^{-2}+s^{-2}\,\ln s +\dots
]\nonumber\\ 
&+&
O(\alpha^6),\ea
\ba (6/\alpha)\,s^2\,\tilde F\,\tilde f 
&=&
(s^{-2}-3+2\,s) \nonumber \\
&+&
(\alpha^2/12)\,[(351/15)\,s^{-2}\nonumber\\
&+&12\,s^{-2}\,\ln s-36\,s^{-1}+40\,s-45\,s^2\nonumber\\
&+&(108/5)\,s^3-4\,s^4]+O(\alpha^4).\ea

Let us denote with $\psi^*$  the transposed complex conjugate wave functions 
associated with  functions $F$ and $G$, and let $\psi$  be the wave functions associated
with  $f$ and $g$. For example, the time component of the 4--vector $
\psi^*\,\gamma_\mu\,\psi$, namely \,$ \psi^*\,\gamma_1\,\psi$\, is
\ba
&{}&\frac{1}{3}|Y_{1,0}|^2\, F\,f+\frac{2}{3}
|Y_{1,1}|^2\,Ff+Y_{0,0}^2\,G g \nonumber\\&=& Y_{0,0}^2\,(F\,f+G\,g),
\ea
 the expression vanishes in a smooth manner at $s=1$. However, graphical discontinuities of the wave
functions are not acceptable. This problem is solved by replacing potential 
$e\,A_\mu$ in favor of $(m\,c^2)\,u_\mu =m\,c^2\,(1,\vec{0})$ in the region beyond the
classical electron radius. The corresponding radial equations are
\bse\ba
\partial_r\,F+\frac{2\,F}{r}-\frac{\beta\,F}{m\,c^2\,r^2} &=& \ \frac{E-m\,c^2}{\hbar
\, c}\,G,\\
-\partial_r\,G+\frac{\beta\,G}{m\,c^2\,r^2} &=& \ \frac{E-m\,c^2}{\hbar
\, c}\,F.\ea\ese 
Now we must set  $E=m\,c^2 $. With this consideration
the external solutions,
\ba G=\exp(-\beta\,s^{-1}),\qquad F=0,\\
f=s^{-2}\,\exp(-\beta\,s^{-1}),\qquad g=0,\ea
and the internal solutions (25)--(26) and (31)--(32) make it possible for the overall
 solution to be  smoothly contiuous in space. The external solutions have no further meaning 
since the product of the two solutions is null (see figure 1).

\section{The densities}

The $\gamma_{\mu}$ matrices (eq. (17)) together with solutions 
(19) and (20) enable one to construct but two entities
with covariant transformation properties\cite{[3]}. 
One of them is a 4-vector
and the other is a pseudo 4-vector, equivalent to a completely antisymmetric 
tensor of rank 3. We shall associate  the 4-vector
with the energy momentum  and the pseudo 4-vector
with spin. The  integrals of these densities are considered to be proportional
 to the value of the physical property in question. In order to write down covariant densities corresponding 
to the recipient charge and to the magnetic moment,   
the Dirac $\gamma_5$ matrix 
appearing in eq. (1) ($\gamma_5 =Diag(1,1,-1,-1)$) is necessary.
With the $\gamma_5$ matrix it is also possible to construct an invariant, 
a pseudo invariant, and an antisymmetric tensor of rank 2. The 
 invariant goes with the recipient charge,  
 the  
pseudo invariant has no particular interpretation, and the 
 antisymmetric tensor goes with the electromagnetic 
polarization . Table (1) shows the 16 densities (4+4+1+1+6) 
of the 5 covariant entities just mentioned. The product  
$\psi_1^* \psi_1$  is denoted as $1\cdot 1$, the product $\psi_2^* \psi_1$ 
as $1\cdot 2$   and so forth.  
Table (2) and (3) show the explicit form of the 4 densities 
that survive volume integration: the time component of the 4-vector, 
the $z$-component of spin, the invariant, and the $z$-component 
of the magnetization 3-vector in the polarization tensor. 
The components $GgY_{0,0}^2$ and $LlY_{0,0}^2$ are missing in tables II and III
because their radial integral can vanish with the proper choice of 
$\alpha$ and $\beta$. All the other densities vanish because they contain products 
of different spherical harmonics . It is relevant 
that the difference between solutions (19) and (20) shows up only 
in those densities involving the $\gamma_5$ matrix. Indeed,
the electron and the positron have opposite electric charge and opposite magnetic moment. the charge 
and  the energy densities have spherical symmetry, spin and 
magnetic moment densities have azimuth symmetry.

\begin{table}
\begin{center}
\caption{}
\begin{tabular}{|c| c| c|} 
\hline 
{Name} &{Operator} &{Density}
\\  
\hline
{} &{} &{}
\\
{Energy} 
&{$ \gamma_{_1}$}
&{$1\cdot 1+2\cdot 2+3\cdot 3+4\cdot 4 $}
\\
{} &{} &{} 
\\
{} 
&{$ \gamma_{_2}$}
&{$1\cdot 4+2\cdot 3+3\cdot 2+4\cdot 1 $}
\\
{} &{} &{}
\\
{} 
&{$\gamma_{_3}$} 
&{$i(-1\cdot 4+2\cdot 3 -3\cdot 2+4\cdot 1 )$}
\\
{} &{} &{}
\\
{}
&{$\gamma_{_4}$} 
&{$1\cdot 3-2\cdot 4+3\cdot 1-4\cdot 2 $}
\\
{} &{} &{}
\\     
\hline
{} &{} &{}
\\
{} 
&{$\,\gamma_{_2}\gamma_{_3}\gamma_{_4}$}
&{$i(1\cdot 3+2\cdot 4+3\cdot 1+4\cdot 2) $}
\\
{} &{} &{} 
\\
{}
&{$i\,\gamma_{_3}\gamma_{_4} $}
&{$-1\cdot 2-2\cdot 1-3\cdot 4-4\cdot 3 $}
\\
{} &{} &{}
\\
{}
&{$i\,\gamma_{_4}\gamma_{_2} $}
&{$i(1\cdot 2-2\cdot 1+3\cdot 4-4\cdot 3) $}
\\ 
{} &{} &{}
\\
{$S_z$}
&{$i\,\gamma_{_2}\gamma_{_3} $}
&{$-1\cdot 1+2\cdot 2-3\cdot 3+4\cdot 4 $}
\\
{} &{} &{}
\\  
\hline 
{} &{} &{}
\\
{Charge $e$}
&{$\gamma_{_5}$}
&{$1\cdot 1+2\cdot 2-3\cdot 3-4\cdot 4 $}
\\
{} &{} &{}
\\
\hline
{} &{} &{}
\\
{}
&{$\gamma_{_2}\gamma_{_3}\gamma_{_4}\gamma_{_5} $}
&{$i(-1\cdot 3-2\cdot 4+3\cdot 4+4\cdot 2) $}
\\
{} &{} &{} 
\\
\hline
{} &{} &{}
\\
{}
&{$i\,\gamma_{_3}\gamma_{_4}\gamma_{_5} $}
&{$-1\cdot 2-2\cdot 1+3\cdot 4+4\cdot 3 $}
\\
{} &{} &{}
\\ 
{}
&{$i\,\gamma_{_4}\gamma_{_2}\gamma_{_5} $}
&{$i(1\cdot 2-2\cdot 1-3\cdot 4+4\cdot 3) $}
\\
{} &{} &{}
\\
{$M_z$}
&{$i\,\gamma_{_2}\gamma_{_3}\gamma_{_5} $}
&{$-1\cdot 1+2\cdot 2+3\cdot 3-4\cdot 4 $}
\\
{} &{} &{}
\\
{}
&{$i\,\gamma_{_2}\gamma_{_5} $}
&{$i(-1\cdot 4-2\cdot 3+3\cdot 2+4\cdot 1) $}
\\
{} &{} &{}
\\ 
{}
&{$i\,\gamma_{_3}\gamma_{_5} $}
&{$-1\cdot 4+2\cdot 3+3\cdot 2-4\cdot 1 $}
\\
{} &{} &{}
\\ 
{} 
&{$i\,\gamma_{_4}\gamma_{_5}$}
&{$i(-1\cdot 3+2\cdot 4+3\cdot 1-4\cdot 2) $}
\\
{} &{} &{}
\\
\hline 
\end{tabular}
\end{center}
\end{table}

\begin{table}
\begin{center}
\caption{Electron}
\begin{tabular}{|c| c| c|}
\hline
{} &{} &{}
\\ 
{} &{$m=1/2$} &{$m=-1/2$}
\\
{} &{} &{}
\\  
\hline
{} &{} &{}
\\
{$E$} 
&{$\frac {1}{3}Ff|Y_{1,0}|^2+\frac{2}{3}Ff|Y_{1,1}|^2$}
&{$\frac{1}{3}Ff|Y_{1,0}|^2+\frac{2}{3}Ff|Y_{1,-1}|^2$}
\\
{} &{} &{} 
\\
\hline
{} &{} &{}
\\
{$S_z$} 
&{$-\frac{1}{3}Ff|Y_{1,0}|^2+\frac{2}{3}Ff|Y_{1,1}|^2$}
&{$\frac{1}{3}Ff|Y_{1,0}|^2-\frac{2}{3}Ff|Y_{1,-1}|^2$}
\\
{} &{} &{} 
\\
\hline
{} &{} &{}
\\
{$e$} 
&{$\frac{1}{3}Ff|Y_{1,0}|^2+\frac{2}{3}Ff|Y_{1,1}|^2$} 
&{$\frac{1}{3}Ff|Y_{1,0}|^2+\frac{2}{3}Ff|Y_{1,-1}|^2$}
\\
{} &{} &{}
\\
\hline
{} &{} &{}
\\     
{$M_z$} 
&{$-\frac{1}{3}Ff|Y_{1,0}|^2+\frac{2}{3}Ff|Y_{1,1}|^2$}
&{$\frac{1}{3}Ff|Y_{1,0}|^2  -\frac{2}{3}Ff|Y_{1,-1}|^2$}
\\
{} &{} &{}
\\ 
\hline 
\end{tabular}
\end{center}
\end{table}

\begin{table}
\begin{center}
\caption{Positron}
\begin{tabular}{|c| c| c|}
\hline
{} &{} &{}
\\ 
{} &{$m=1/2$} &{$m=-1/2$}
\\
{} &{} &{}
\\  
\hline
{} &{} &{}
\\
{$E$} 
&{$\frac{1}{3}Kk|Y_{1,0}|^2+\frac{2}{3}Kk|Y_{1,1}|^2$}
&{$\frac{1}{3}Kk|Y_{1,0}|^2+\frac{2}{3}Kk|Y_{1,-1}|^2$}
\\
{} &{} &{} 
\\
\hline
{} &{} &{}
\\
{$S_z$} 
&{$-\frac{1}{3}Kk|Y_{1,0}|^2+\frac{2}{3}Kk|Y_{1,1}|^2$}
&{$\frac{1}{3}Kk|Y_{1,0}|^2-\frac{2}{3}Kk|Y_{1,-1}|^2$}
\\
{} &{} &{} 
\\
\hline
{} &{} &{}
\\
{$e$} 
&{$-\frac{1}{3}Kk|Y_{1,0}|^2-\frac{2}{3}Kk|Y_{1,1}|^2$} 
&{$-\frac{1}{3}Kk|Y_{1,0}|^2-\frac{2}{3}Kk|Y_{1,-1}|^2$}
\\
{} &{} &{}
\\
\hline
{} &{} &{}
\\     
{$M_z$} 
&{$\frac{1}{3}Kk|Y_{1,0}|^2-\frac{2}{3}Kk|Y_{1,1}|^2$}
&{$-\frac{1}{3}Kk|Y_{1,0}|^2  +\frac{2}{3}Kk|Y_{1,-1}|^2$}
\\
{} &{} &{}
\\ 
\hline 
\end{tabular}
\end{center}
\end{table}

\section{Parameters $\alpha$ and $\beta$}
The components $1/3Ff|Y_{1,0}|^2$ and $2/3Ff|Y_{1,1}|^2$ could be 
named lepton-quarks, two subparticles revolving around the singular
point of the potentials and carrying the physical properties of the 
electron. Mathematically, it is equivalent to the vanishing of the integral 
of the component $GgY_{0,0}^2$, which component is by itself the 
first of the two elementary invariant densities:
	\ba
	2\pi\,(u_\mu \psi^*\gamma_\mu \psi- \psi^*\gamma_5 \psi) &=&
	Gg \ \equiv \ I_1, \\
	2\pi \,(u_\mu \psi^*\gamma_\mu \psi+ \psi^*\gamma_5 \psi) &=&
	Ff \ \equiv \ I_2, 
	\ea
where $u_\mu$ is the 4-velocity of the singularity. Thus, the first equation instrumental for the
determination of $\alpha$ and $\beta$ is
	\begin{equation}
	\int_{space} I_1 =0.
	\end{equation}
Now, if $\eta\,=\,2\beta\ \,\ll\,1$ and $w=\exp (-\eta s^{-1})$, the 
following approximations hold good.
	\ba
	\int^1_0 ws^{-3} ds &=& e^{-\eta}(\eta^{-2} + \eta^{-1})\approx
	\eta^{-2},\\
	\int^1_0 ws^{-2} ds &=& e^{-\eta} \eta^{-1} \approx \eta^{-1},\\
	\partial_{\eta}\int^1_0 ws^{-1} ds &=& -\int_0^1 ws^{-2}ds,
	\ea
so, 
	\begin{equation}
	\int^1_0 ws^{-1} ds \ \approx \ -\ln \eta -0.577216664906...\,,
	\end{equation}
where the constant of the integration is the Euler-Mascheroni constant.

Iterative differentiation with respect to $\eta$ gives 
	\ba
	\int^1_0 w ds       &\approx& 1+\eta \,\ln (\eta), \\
	\int^1_0 w s^{k} ds &\approx& \frac {1}{k+1} -\frac {1}{k} \eta
	\qquad {\rm for} ~~ k\ge 1.
	\ea

From eqs. (23) and (36) it is easy to realize that eq. (44) is satisfied 
if $\eta$ is much smaller than $\alpha$. Further, from the integrals 
just shown, it follows that eq. (44) and the resulting relation between 
$\alpha$ and $\beta$ are, in very good approximation,
	\ba
	\int_0^1 w\,[ \,(1-s)^{2} -\frac {\alpha^2}{12}s^{-2}\,]\,ds &=& 0,\\
	\beta &=& \frac {\alpha^2}{8}.
	\ea

Since all the elements of the theory should play their own role,
the second  condition necessary for the determination of 
$\alpha$ and $\beta$ should rely on $I_2$. We need to postulate an underlying relation
 between the conceptual structure of QM with pure mathematics. 
Specifically, invariant $I_2$ is regarded as the fundamental 
eigendensity satisfying,
	\begin{equation}
	\int_{\textrm{space}} \Lambda \,I_2 =\int_{\textrm{space}} \lambda \,I_2,
	\end{equation}
where $\Lambda$ is function of the invariants associated 
with the interaction 
in eq. (18),
	\begin{equation}
	\Lambda=\Lambda\,( \,[A_\mu A_\mu]^{1/2}, [B_\mu B_\mu]^{1/2}\,)
	=\Lambda\big(\alpha s^{-1}, \beta s^{-2}\big).
	\end{equation}
If invariant $I_2$ is interpreted as the probability density
of the recipient  charge, then eq. (53) is a statement about the 
expected value of $\Lambda$. The eigenvalue $\lambda$ is to be 
determined independently of $\Lambda$ through a 
mathematical criterion forbidding the use 
of numerical coefficients and based on proportion and distinction, 
as follows: consider the integral
	\begin{equation}
	\int_0^1 \left(\frac{\alpha}{s}\right) (s^{-2} -3 +2s)\,wds,
	\end{equation}
which is the integral of $(A_\mu A_\mu)^{1/2}I_2$ neglecting terms 
in $\alpha^2$ (see eq. (37)). The contribution of the first term is much larger 
than the contribution of the other two terms. 
Now, consider integrals with the invariants inverted
	\ba
	&{}&\int_0^1 \left(\frac{s}{\alpha}\right) (s^{-2} -3 +2s)\,wds\nonumber\\
	&\approx& \frac{1}{\alpha}\,\left[ -\ln(2\beta) -0.577217 -\frac{3}{2}
	+\frac{2}{3}\,\right],
	\ea
	\begin{equation}
	\int_0^1{\left(\frac{s^2}{\beta}\right) (s^{-2} -3 +2s)\,wds}
	\ \approx \ \frac{1}{2\beta}.
	\end{equation} 

In this case the three terms are significant to the value 
of the integrals (56) and (57). Therefore, the assumption will be made that eq. (53) has the form,
	\ba
	&&\int_{space}  \big[ \,\Omega_1 (A_\mu A_\mu)^{-1/2}
	-\Omega_2 (B_\mu B_\mu)^{-1/2}\,\big] \,I_2 \nonumber \ = \\
	&&\int_{space}
	\lambda(\alpha,\beta) \,I_2,
	\ea
where $\Omega_1$, $\Omega_2$, $\lambda$ are as yet undefined functions 
of $\alpha$ and $\beta$.

The right hand side of eq. (58) becomes
\begin{equation}
	\lambda \int_0^1{ (s^{-2} -3 +2s)\,wds}
	\ \approx \ \frac{\lambda}{2\beta}.
	\end{equation} 
However, integral (57) is much larger than the integral (56).
The weight of the three terms under integral signs in eq (58) would have the same 
order of magnitude if $\Omega_2\approx \alpha$,\, $\Omega_1\approx 1$ and $\lambda\approx \alpha$.

We assume also a functional relation,
	\begin{equation}
	\Omega_1 -\Omega_2=1,
	\end{equation}
and in order to treat $\alpha$ and $\beta$ on a par, we take
	\ba
	\Omega_1 \ = \ \cosh^2 (\alpha -\beta)^{1/2} &\approx& 1, \\
	\Omega_2 \ = \ \sinh^2 (\alpha -\beta)^{1/2} &\approx& \alpha -\beta.
	\ea

The eigenvalue $\lambda$ is considered function of
$\alpha -\beta$. To have a sharp distinction between $(\Omega_1, \Omega_2)$
and $\lambda$, we express $\lambda$ not as an infinite sum 
but as the product of infinite factors rapidly converging on $1$.
The exponent of the factors being increasing powers of the dimensionless
unit of charge $\alpha^{1/2}$. Thus, the explicit form of eq. (58) becomes 
	\ba
&&	\int_{\textrm{space}}
	\big[\cosh^2 (\alpha -\beta)^{1/2}~~(A_\mu A_\mu)^{-1/2}-\nonumber\\
            &&\qquad \quad\sinh^2 (\alpha -\beta)^{1/2}~~(B_\mu B_\mu)^{-1/2}\big]\,
	    I_2 \nonumber \ =\nonumber\\
\nonumber\\
	    &&\int_{\textrm{space}} 
	I_2\,\frac {(\alpha-\beta)^{\alpha^0} (\alpha-\beta)^{\alpha^1}
	 (\alpha-\beta)^{\alpha^2} \cdots}
	{(\alpha-\beta)^{\alpha^{1/2}} (\alpha-\beta)^{\alpha^{3/2}}
	 (\alpha-\beta)^{\alpha^{5/2}} \cdots}\nonumber\\
	\ea

Eq. (63) unveils the inextricable link between physical reality and the elegance of mathematical
symmetry. Substituting the value of $\beta$ as given in (52) and using the 
approximations (56), (57) and (59) we get
\ba &-&\cosh^2([\alpha-\alpha^2/8]^{1/2})\,(1/\alpha)\,[\ln\,(\alpha^2/4)+1.41055]\nonumber\\
&-&\sinh^2([\alpha-\alpha^2/8]^{1/2})\,(4/\alpha^2)\nonumber\\&=& \ (4/\alpha^2)\,
(\alpha-\alpha^2/8)^{1/(1+\sqrt{\alpha})}.
\ea

Eq. (53)
is satisfied for $\alpha$ very close to $0.007292$. The contribution of all the 
terms as given by eqs. (36) and (37) and the consideration of the 
weight of all terms missing to get the exact expressions of $g G$ and $Ff$ only slightly  shifts 
the former value to $\alpha=0.0072976(3)$. This value is within the range of experimental
precission, but further numerical analysis is necessary to determine if it coincides with the 
experimental value $\alpha= 0.007297357(2)$.   
 
In conclusion, the electron is the physical 
realization of relativity, which almost a 
century ago put a definite end to the idea of the ether. 
Here again relativity shows  that space has 
no physical properties: the philosophy to which the negative 
energy of the Dirac positron gave rise, 
and the Heisenberg uncertainty relations 
are consequence of a deficiency inherent to the 
costumary wave equations: selfaction is missing.

\section*{Acknowledgments}
This work was finished at the Instituto 
de  Ciencias Nucleares,  Universidad Nacional
Aut\'onoma de M\'exico (UNAM). I enjoyed interesting conversations
 with Drs. A. Frank, O. Casta\~nos, G. Monsivais and R. Sussman.


\begin{thebibliography}{99}

\bibitem{[1]} J. Schwinger, \textit{Phys Rev}, \textbf{76}, 817, (1948).  
\bibitem{[2]} R.P. Feynman, \textit{Phys Rev}, \textbf{76}, 776, (1949).  

\bibitem{[3]} Hill E.L.   \textit{Rev. Mod. Phys.} \textbf{10}, 2, pp. 87-118,
(1938). 



\end{thebibliography}
\end{document}